\documentclass[9pt,twocolumn,twoside]{optica}
\setboolean{shortarticle}{false}
\setboolean{minireview}{false}

\title{Reliable coherent optical memory based on a laser-written waveguide}

\author[1,2]{Chao Liu}
\author[1,2,*]{Zong-Quan Zhou}
\author[1,2]{Tian-Xiang Zhu}
\author[1,2]{Liang Zheng}
\author[1,2]{Ming Jin}
\author[1,2]{Xiao Liu}
\author[1,2]{Pei-Yun Li}
\author[1,2]{Jian-Yin Huang}
\author[1,2]{Yu Ma}
\author[1,2]{Tao Tu}
\author[1,2]{Tian-Shu Yang}
\author[1,2,$\dag$]{Chuan-Feng Li}
\author[1,2]{Guang-Can Guo}

\affil[1]{CAS Key Laboratory of Quantum Information, University of Science and Technology of China, Hefei 230026, People's Republic of China}
\affil[2]{CAS Center For Excellence in Quantum Information and Quantum Physics, University of Science and Technology of China, Hefei 230026, People's Republic of China}

\affil[*]{Corresponding author: zq\_zhou@ustc.edu.cn}
\affil[$\dag$]{Corresponding author: cfli@ustc.edu.cn}

\doi{\url{https://doi.org/10.1364/OPTICA.379166}}

\begin{abstract}
{$\mathrm {^{151}Eu^{3+}}$-doped yttrium silicate ($\mathrm {^{151}Eu^{3+}:Y_2SiO_5}$ ) crystal is a unique material that possesses hyperfine states with coherence time up to 6 h. Many efforts have been devoted to the development of this material as optical quantum memories based on the bulk crystals, but integrable structures (such as optical waveguides) that can promote $\mathrm {^{151}Eu^{3+}:Y_2SiO_5}$-based quantum memories to practical applications, have not been demonstrated so far.} Here we report the fabrication of type \uppercase\expandafter{\romannumeral2} waveguides in a $\mathrm {^{151}Eu^{3+}:Y_2SiO_5}$ crystal using femtosecond-laser micromachining. The resulting waveguides are compatible with single-mode fibers and have the smallest insertion loss of $4.95\ dB$. On-demand light storage is demonstrated in a waveguide by employing the spin-wave atomic frequency comb (AFC) scheme and the revival of silenced echo (ROSE) scheme. We implement a series of interference experiments based on these two schemes to characterize the storage fidelity. Interference visibility of the readout pulse is $0.99\pm 0.03$ for the spin-wave AFC scheme and $0.97\pm 0.02$ for the ROSE scheme, demonstrating the reliability of the integrated optical memory.
\end{abstract}

\setboolean{displaycopyright}{true}
\begin{document}

\maketitle
\section{Introduction}
\begin{figure*}[htbp]
\centering
\fbox{\includegraphics[width=0.85\linewidth]{setup.pdf}}
\caption{Schematic of the experimental set up. The main laser beam is split into two to be employed as preparation $\&$ control mode and input mode respectively. The input mode can be injected through input 1 mode in the spin-wave AFC scheme or input 2 mode in the ROSE scheme. The yellow light beams represent optical paths for the spin-wave AFC memory scheme. The red beams denote the input and output mode for ROSE scheme. Labels at the left corner are: Beam Expander (BE), Faraday Rotator (FR), Half Wave Plate (HWP), Fiber Coupler (FC), Beam Splitter (BS), Polarizing Beam splitter (PBS), lens and mirror. }
\label{fig.setup}
\end{figure*}
Optical quantum memories, which can map quantum information between light and matter with high fidelity, are crucial devices in the implementation for large-scale quantum networks\cite{gisin2008nature}\cite{gisin2011rmp}. As a promising candidate for this device, quantum memories with high fidelity \cite{gisin2008nature}\cite{ zzq2012prl}\cite{zzq2015prl}, multi-mode capacity \cite{zzq2015prl, gisin2010nature, zzq2015nc, zzq2018nc} as well as long storage time \cite{afz2015prl}\cite{afz2017prl}, have been demonstrated in bulk rare-earth-ion-doped crystals. To promote this remarkable device to practical application, many efforts have been devoted to the development of integrated quantum memories. Three approaches have been employed for manufacturing integrated quantum memories in rare-earth-ion-doped crystals. The first one is using industry-standard Ti indiffusion in $\mathrm {LiNbO_3}$. High fidelity \cite{tit2007prl}\cite{gisin2007prl}, broadband \cite{tit2017prl}\cite{tit2011nature}, multiplexed \cite{tit2014prl} and telecom-wavelength \cite{tit2019prap} memory as well as integrated processor \cite{tit2014njp} have been demonstrated. The second one is using focused-ion-beam milling. Quantum memories with high fidelity \cite{fara2017sci} and telecom-wavelength \cite{fara2019prap} have also been demonstrated. However, the storage time and storage efficiencies in these two systems are significantly reduced as compared with that in bulk crystals \cite{gisin2010nature}\cite{tm2010}\cite{gisin2010prl}. As a result, spin-wave storage with extended lifetime have not been achieved based on these two approaches so far. The third one is using femtosecond-laser micromachining (FLM). Recently, integrable memories are successfully demonstrated, based on waveguides fabricated in a $\mathrm {Pr^{3+}:Y_2SiO_5}$ crystal using FLM \cite{reid2016prap, reid2018optica, reid2019prl}. Storage time of up to 15 $\mu$s \cite{reid2016prap}, storage efficiency as high as $21\ \%$ \cite{reid2018optica}, and storage modes as many as 130 \cite{reid2019prl}, have been achieved. These results are comparable with the performance of bulk crystals \cite{reid2017prl, reid2010prl, reid2015prl}, thus making FLM an appealing technique to achieve integrated quantum memories. $\mathrm {^{151}Eu^{3+}:Y_2SiO_5}$ crystal is an attractive candidate for optical quantum memories because of the longest storage time for single photons \cite{afz2015prl}\cite{afz2017prl} and the longest spin coherence time \cite{nature_6hour} among the solid-state optical memories. However, there have been no demonstration of integrated memories in this material. More importantly, storage fidelity of the memory after fabrication using FLM has never been quantitatively characterized before, which is a critical figure-of-merit for coherent optical memories \cite{gold2013prl}.

In this paper, we report the fabrication of the so-called type \uppercase\expandafter{\romannumeral2} waveguides \cite{cf2014lpr} in a $\mathrm {^{151}Eu^{3+}:Y_2SiO_5}$ crystal using FLM. Optimized fabrication parameters ensure that the waveguides are compatible with single-mode fibers (SMFs). On-demand light storage, based on the spin-wave atomic frequency comb (AFC) scheme \cite{gisin2008nature}\cite{gisin2009pra} and the revival of silenced echo (ROSE) scheme \cite{rose2011njp}, are demonstrated in the waveguide section, respectively. Then, storage fidelity of the memory after fabrication is characterized by performing a series of interference experiments based on these two schemes.

\section{WAVEGUIDE FABRICATION}
 The substrate used here is an isotopically enriched $\mathrm {^{151}Eu^{3+}:Y_2SiO_5}$ crystal (isotope enrichment of $99.9\ \%$), with a dimension of $15 \times 5 \times 4$ mm $(b \times D_1 \times D_2)$ and an ion concentration of $0.1\ \%$. This crystal has the maximum absorption for the $ ^7F_0$ $\rightarrow $ $^5D_0$ transition at 580 nm when the light is polarized along the D1 axis of the crystal. A FLM system from WOPhotonics (Altechna R$\&$D Ltd, Lithuania) is utilized to execute the fabrication. The femtosecond-laser beam, with a wavelength of 1030 nm, is injected along the $D_2$ axis. It is then focused 150 $\mu$m beneath the top face of the crystal with a $50\times$ objective ($NA=0.65$). Two parallel damage tracks at a distance of 20 $\mu$m are formed by translating the crystal along the b axis with the speed of 575 $\mu$m/s when the femtosecond-laser is shining. The fabrication parameters are: pulse duration of 300 fs, energy per pulse of 931 nJ, repetition rate of 201.9 kHz. We notice that these optimized fabrication parameters are largely different from that which are used to fabricate the same type of waveguides in $\mathrm {Pr^{3+}:Y_2SiO_5}$ crystals \cite{reid2016prap}\cite{reid2018optica}. 

 The laser beam with wavelength of 580 nm can be effectively confined between these two tracks only when it is in horizontal polarization (parallel to the $D_1$ axis). {This is compatible with the requirement of the efficient absorption for light in $\mathrm {Eu^{3+}:Y_2SiO_5}$ crystals.}
 Single-mode coupling efficiency of the laser beam when it's in horizontal polarization is typically 2 orders higher than that of in vertical polarization. This is consistent with previous work since type \uppercase\expandafter{\romannumeral2} waveguides usually support only one polarization mode \cite{reid2016prap}\cite{reid2018optica}\cite{apbtype2}. We fabricate several waveguides in the same crystal and choose the most efficient one to perform the experiments, as different waveguides may have diverse coupling efficiencies even with the same fabrication parameters, owing to defects in the crystal.

 \section{ EXPERIMENTAL SETUP}
The laser source for the optical memory experiment is a frequency-doubled semiconductor laser (TA-SHG, toptica), which has an output power of 800 mW at a frequency of 516.848 THz. A high-stable cavity placed in a high-vacuum chamber is employed to lock the laser frequency, resulting in a line-width of sub-kHz. A cryostat (Montana Instrument) with the lowest working temperature of approximately 3.2 K is utilized to cool the sample. The sample is mounted on a 3-axis translator which has a positioning accuracy of 10 nm. All experiments are performed at low vibration time windows after synchronizing with the cycle of the cold head, which has a vibration period of approximately 700 ms. {The total usable period for experiments is approximately 250 ms.} As shown in Fig. \ref{fig.setup}, acoustic optic modulators (AOM) are employed to modulate the optical pulses. AOM 1 and AOM 2 are in double-pass configuration and are used to control the amplitude and frequency of each pulse. AOM 3 is in single-pass configuration and serves as an optical switch. Specifically, AOM 1 and AOM 3 are driven by an 8-channel arbitrary waveform generator (HDAWG, Zurich Instruments). Meanwhile, AOM 2 is driven by a modulated RF source, which can coherently control the phases of the optical pulses for heterodyne detections.

\begin{figure}[htbp]
\centering
\fbox{\includegraphics[width=0.98\linewidth]
{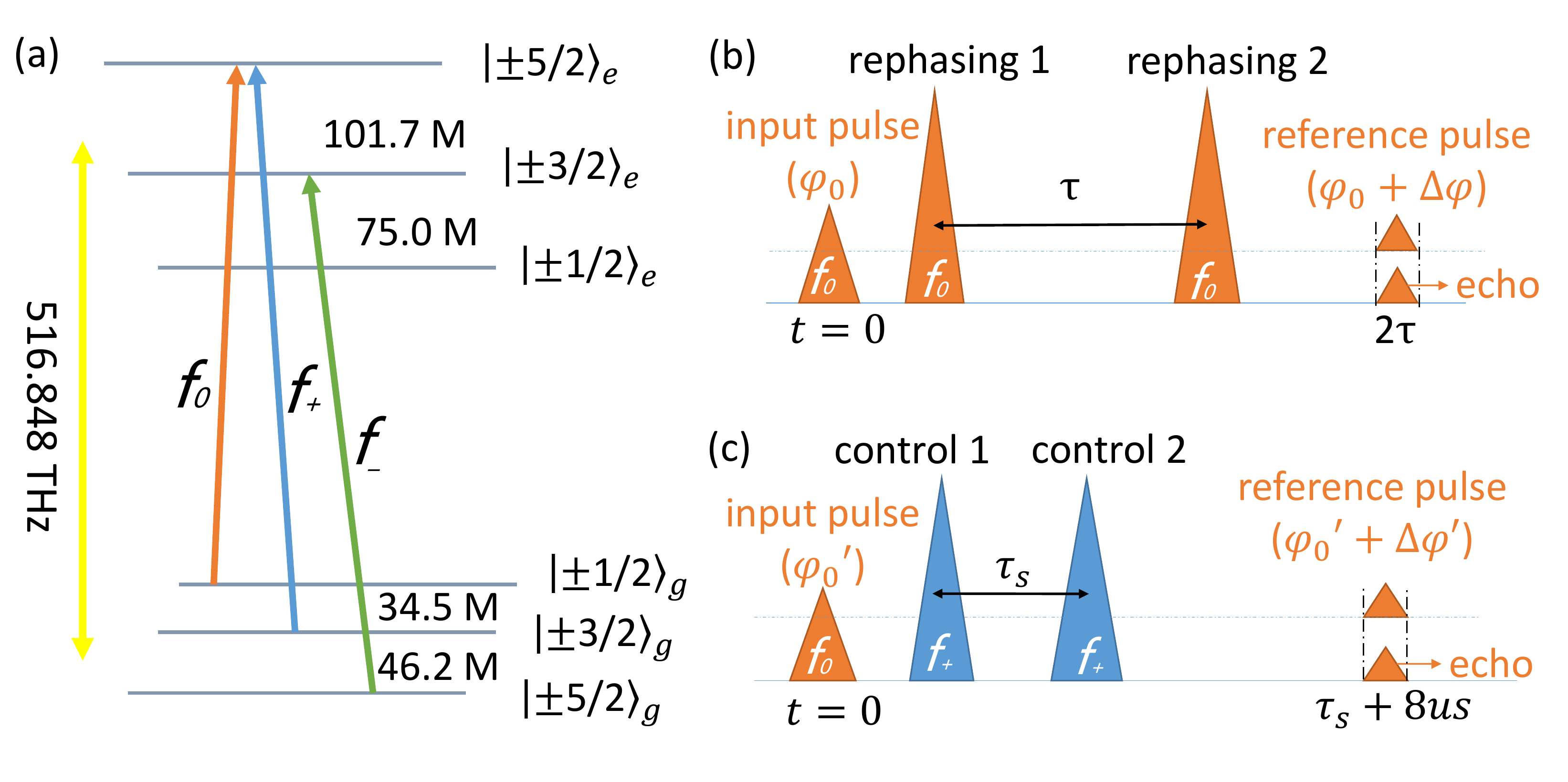}}
\caption{(a) Energy level diagram of the $ ^7F_0$ $\rightarrow $ $^5D_0$ transition of $\mathrm {^{151}Eu^{3+}:Y_2SiO_5}$ at zero magnetic field. $f_0$ (orange), $f_+$ (blue) and $f_-$ (green) denote optical transitions with different hyperfine levels. (b) Time sequence for the ROSE interference experiment. (c) Time sequence for the spin-wave AFC interference experiment. The triangles represent the optical pulses, while their frequencies are distinguished by the filled colors. The reference pulses are injected when the echoes come out. {In the interference experiments, the time delays are fixed as $\tau =4.94$ $\mu$s for ROSE scheme and $\tau_s = 1.7$ $\mu$s for spin-wave AFC scheme respectively.} }

\label{fig.energy_level}
\end{figure}

For the spin-wave AFC scheme, input 1 and preparation $\&$ control beam are combined by a 92:8 beam splitter (BS). The combined light beam is expanded from a diameter of 2 mm to 6 mm by a beam expander (BE) and then focused at one end of the waveguide using a plano-convex lens ($f=75$ mm). This lens is also mounted on a 3-axis translator which has longer traveling distances. {The focal spot has a diameter of $D=5.4\ \mu m$. This is calculated using the Gaussian beam focussing formula $D \approx \frac{2.36f\lambda}{\pi D_0}$. Here $D_0$ (6 mm) is the beam diameter before the plano-convex lens. Under the help of a long working distance microscope above the cryostat, strong red fluorescence can be observed when the laser beam passes through the crystal and resonates with the $\mathrm {^{151}Eu^{3+}}$ ions. By carefully adjusting the two 3-axis stage, the red fluorescence can be almost completely confined between the two damage tracks.} Output laser beam after the waveguide is collected by a fiber coupler (FC 7) after three lenses of different focal lengths, and then is sent to a photo detector or a photo multiplier tube (PMT) after passing the optical switch (AOM 3).

The optimized coupling efficiency into the SMF is $24\ \%$, which is calculated as the ratio between the power of the output beam after FC 7 and input power before the cryostat. Considering the transmission efficiency of $75\ \%$ for other optical elements, insertion loss of the waveguide can be calculated as $4.95\ dB$. We notice that the insertion loss here is approximately half of that has been achieved in industry-standard waveguides \cite{tit2017prl}\cite{tit2011nature}\cite{tit2019prap}. For the ROSE scheme, the preparation $\&$ control mode is the same as the spin-wave AFC scheme. However, the input 2 mode is injected from FC 5 and then collected by FC 6. Thus the input mode and control mode are in counter-propagating configuration, to silent the standard photon echo using phase mismatching \cite{rose2011njp}.

\section{RESULTS}

\subsection{SAMPLE CHARACTERIZATION}
Spectral preparation is the prerequisite for the implementation of optical memory. This is achieved by two steps based on spectral hole burning technique \cite{afz2012prb_specinvest}. The first step named class cleaning. Preparation pulses with center frequency of $f_0$, $f_+$($=f_0+34.5\ MHz$) and $f_-$($=f_0-20.9\ MHz$) are applied simultaneously. The preparation pulses last for 2 ms and repeat for 100 times. After this process, one class ions is picked up and other classes ions are all pumped away. The second step is called spin polarization. This step is done by applying preparation pulses with center frequency of $f_+$ and $f_-$ and repeating them for 100 times. During these two steps, each frequency is swept over 4 MHz. Finally, the ions are polarized to ${\left | \pm {1/2} \right \rangle}_g$ over this frequency range.

\begin{figure}[htbp]
\centering
\fbox{\includegraphics[width=0.95\linewidth]
{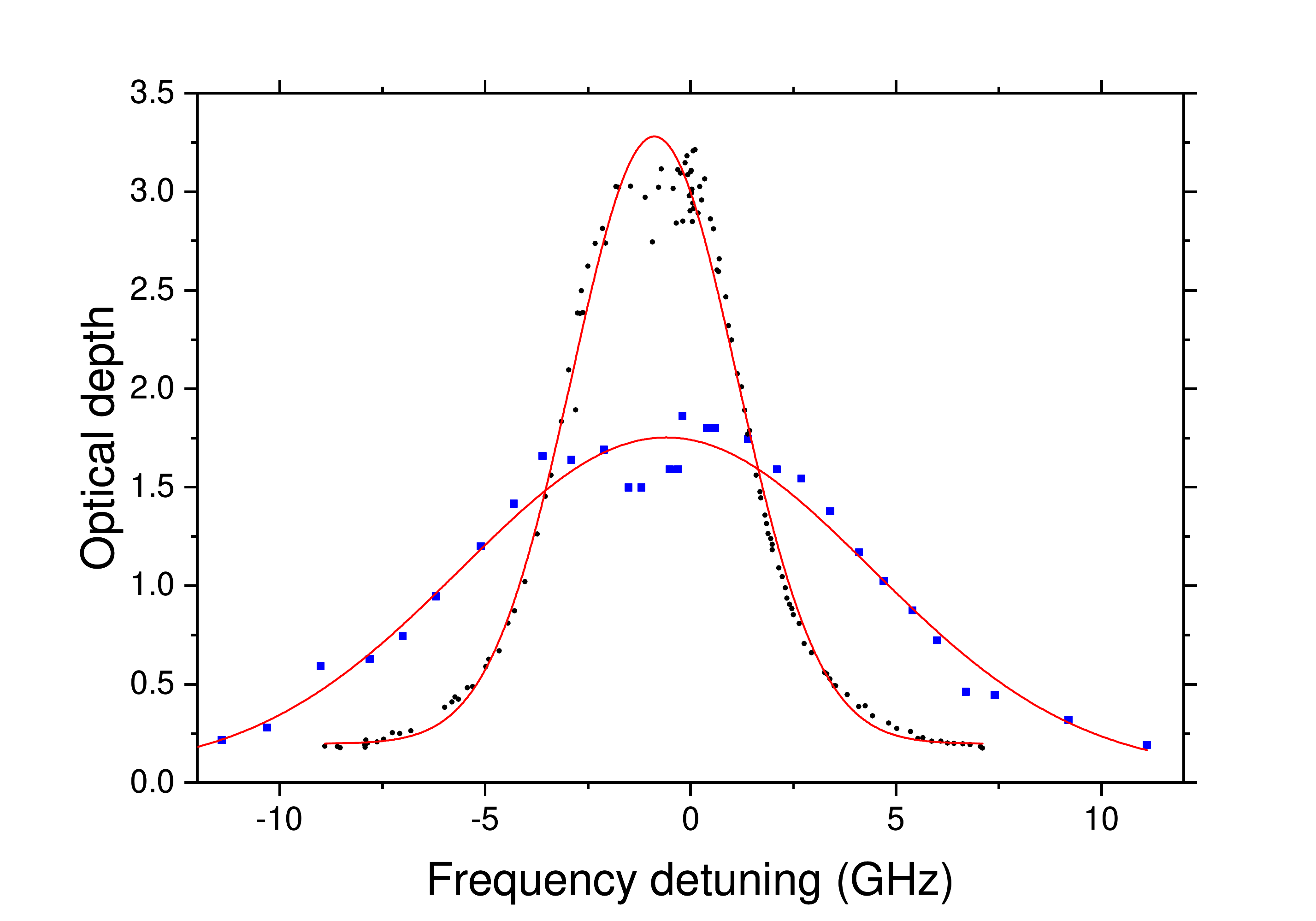}}
\caption{Absorption profile of the sample. Black dots are experimental data collected in the bulk section, while the experimental data in the waveguide are represented in blue squares. Gaussian fitting gives the optical inhomegeneous broadening $\Gamma_{inh} (FWHM)$: 4.7 GHz in the bulk section and 11.8 GHz in the waveguide. Zero detuning corresponds to 516.848 THz.}
\label{fig.OD}
\end{figure}
{The input pulse is a Gaussian pulse with frequency of $f_0$ and full width at half maximum (FWHM) duration of approximately 500 ns. The intensity ($P_a$) of the transmission pulse is recorded by a photon detector (Thorlabs, PDA8A). The intensity of the input pulse ($P_0$) is obtained by burning a transparent window at $f_0$. In the waveguide section, $P_a/P_0$ is measured to be $17.4\ \%$. It corresponds to an OD of 1.75. In contrast, an OD of 3.00 is measured in the bulk crystal. The reduction of OD can be explained by the expansion of the optical inhomogeneous broadening in the waveguide after fabrication, which is broadened by more than 2 times compared with that of in the bulk section (see fig. \ref{fig.OD}).}

\begin{figure}[htbp]
\centering
\fbox{\includegraphics[width=0.95\linewidth]
{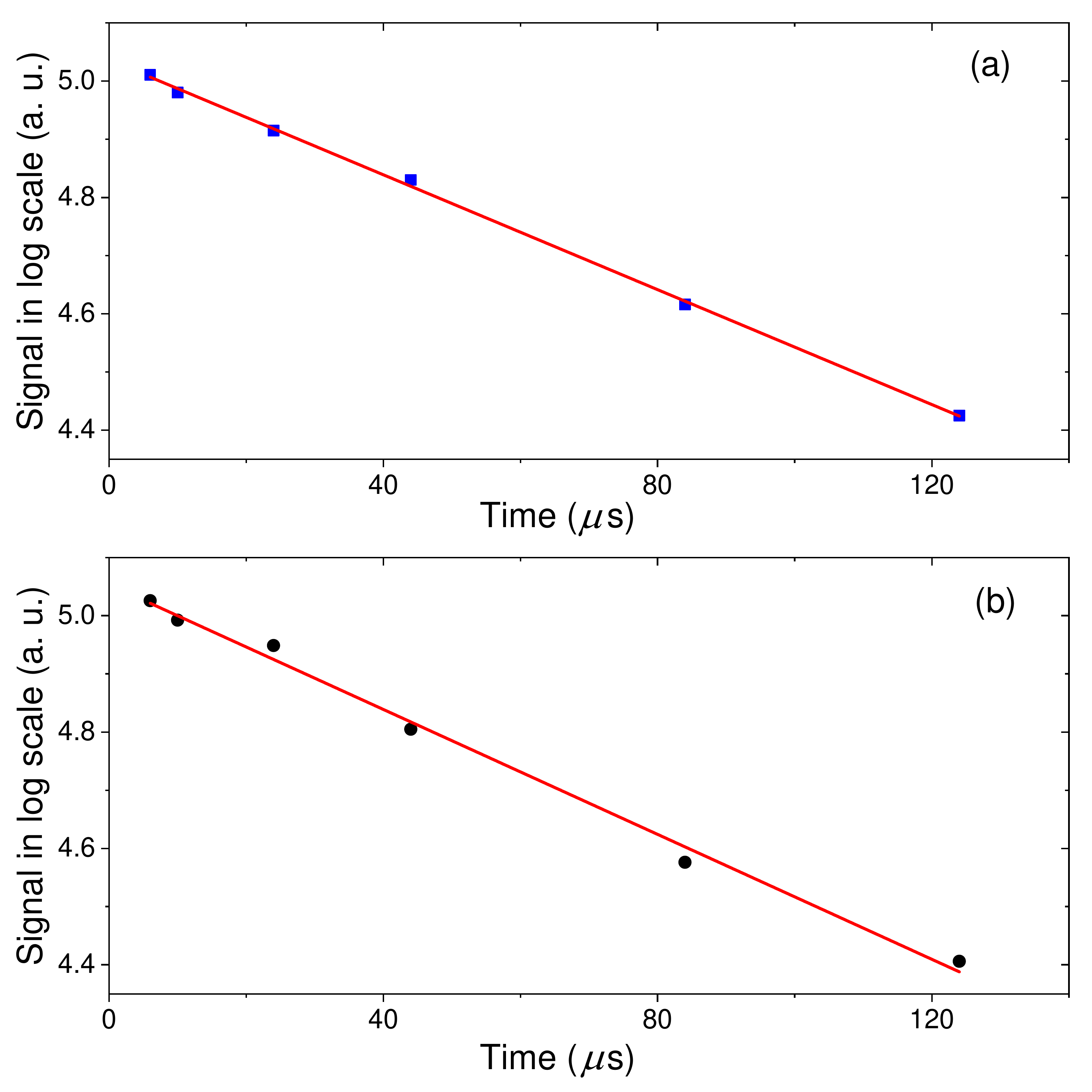}}
\caption{Photon echo amplitude (in logarithmic scale) as a function of time spacing between the two pulses. (a) Measurement performed in the waveguide section. Linear fitting gives $T_2=202 \pm {3}$ us.  (b) Measurement performed in the bulk section. Linear fitting gives $T_2=186 \pm{7}$ us. }
\label{fig.T2}
\end{figure}

{The optical coherence time $T_2$ is measured by performing a series of two-pulse photon echo experiments. The first input pulse, with a duration (FWHM) of 500 ns, is injected at $t=0$, generating a superposition state between ${\left | \pm {1/2} \right \rangle}_g$ and ${\left | \pm {5/2} \right \rangle}_e$. Then, a refocusing pulse is injected at $t=\tau_1$ and a photon echo is induced at $t=2\tau_1$. Due to smaller amount ions involved during the whole process, the photon echo is small and is not easy to be detected by a photon detector when $\tau_1$ is larger than 10 $\mu$s. So a heterodyne detection is employed here to obtain the echo amplitude. $T_2$ is extracted from the exponential decay of echo amplitude as $\tau_1$ increases (shown in Fig. \ref{fig.T2}). The peak power of the input pulse is 0.5 mW in the waveguide section. The fitted $T_2=202\pm {3}\ \mu$s. As a comparison, the peak power is approximately 5.3 mW in the bulk section. The fitted $T_2=186\pm{7}\ \mu$s . This result indicates that the coherence property of the memory is not affected during the fabrication process.}

\subsection{ROSE STORAGE}
The ROSE scheme is free from the noise induced by the inversion of medium population in the conventional two-pulse photon echo scheme \cite{pe1964prl}, thus has the potential to demonstrate optical storage at the single-photon level \cite{rose2011njp}\cite{rose2014lp}. In ROSE scheme, a second rephasing pulse is applied to bring the atoms excited by the first rephasing pulse back to the ground state, and to reverse the phases of them, leading to a secondary echo. The primary echo is silenced when spatial phase mismatching is fulfilled such as the counter-propagating configuration used here.

\begin{figure}[htbp]
\centering
\fbox{\includegraphics[width=0.95\linewidth]
{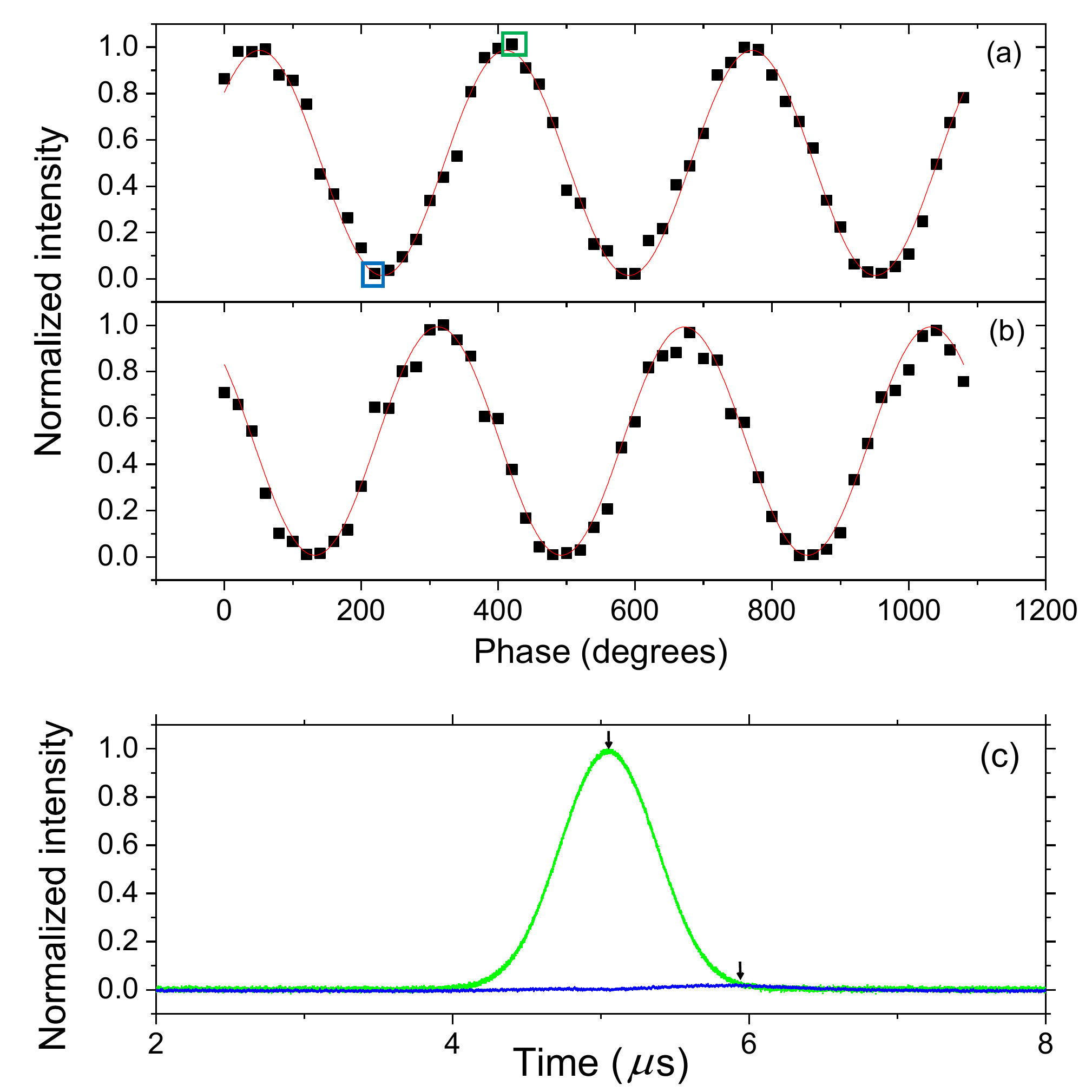}}
\caption{Normalized intensity of the output light as a function of the relative phases between the input pulse and reference pulse. Interference visibilities are fitted by using Eq (\ref{eq1}). (a) For the ROSE scheme, $V=0.97 \pm{0.02}$. (b) For the spin-wave AFC scheme, $V=0.99 \pm{0.03}$. {(c) Examples of interference patterns for ROSE scheme. Blue lines corresponds to data marked with blue square and green lines corresponds to data marked with green square in Fig. \ref{fig.interference}.(a) . Black arrows indicate the peak positions. The peak of the blue curve is away from the original pulse center because of the imperfect interference in the tail of the pulse.}
}
\label{fig.interference}
\end{figure}

To benchmark the memory performance in terms of the coherent storage, here we implement an interference experiment to characterize the storage fidelity \cite{gold2013prl}\cite{fid2010np}\cite{fid2013prl}. The experimental time sequence is shown in Fig. \ref{fig.energy_level} (b). The input pulse is the same as that utilized in the measurements of absorption and $T_2$. Then two rephasing pulses (also at frequency $f_0$) with spacing of $\tau$ are injected, resulting in a ROSE echo at $2\tau$. Complex hyperbolic secant (CHS) pulses \cite{CHS1985pra}\cite{CHS2005pra} are utilized here as the rephasing pulses. They have a duration of 0.94 $\mu$s and a peak power of 13 mW (before the cryostat). A Gaussian reference pulse with relative phase of $\Delta \varphi$ to the first input pulse is injected at $2\tau$, to interfere with the echo. {Careful alignments are made to ensure temporal overlap between the echo and the reference pulse. The output signals are measured by the photon detector and displayed on an oscilloscope. We read the peak value of each interference pattern to form the final interference curve.} Fig. \ref{fig.interference} (a) shows the interference curve with a fixed $\tau$ ($4.94$ $\mu$s) when $\Delta \varphi$ increases 20 degrees step by step. Intensity of the interference signal can be represented by:
\begin{equation}
    I(\varphi)=\frac{I_{max}}{2}\left[1+V\sin{(\varphi+\varphi_1)} \right] ,
\label{eq1}
\end{equation}
where $I_{max}$ is the maximum signal intensity, $\varphi$ is the relative phase between the input pulse and reference pulse, $\varphi_1$ is the phase offset, $V$ is the interference visibility. The interference curve is well fitted by this formula with $V=0.97 \pm{0.02}$, indicating that the relative phases are well preserved during the whole storage process.

Considering the finite transfer efficiency of the rephasing pulses, the storage efficiency can be expressed as \cite{rose2011njp}:

\begin{equation}
    \eta={\eta_T}^2(OD)^2e^{-OD}e^{-4\tau/T_{2eff}} ,
\label{eq2}
\end{equation}
$\eta_T$ denotes the transfer efficiency of a single rephasing pulse, $T_{2eff}$ is the effective coherence time. Fig. \ref{fig.T2eff} (a) shows the storage efficiency as a function of the storage time. The fitted $T_{2eff}=37.4 \pm 0.9 $ $\mu$s, is obviously shorter than the optical coherence time. This is attributed to the severe instantaneous spectral diffusion (ISD) effect \cite{reid2018optica}\cite{ID1962prl}\cite{rose2015njp} induced by the massive rephasing pulses. {A possible way to minimize this effect is reducing the storage bandwidth \cite{rose2015njp}. Moreover, longer storage time can be expected by transferring the coherence to spin states \cite{rose2014pra}.} The maximally possible storage efficiency is extracted as $\eta_0=34.4\ \%$ when $\tau=0$. $\eta_T$ is then deduced as $80\ \%$.

\subsection{SPIN-WAVE AFC STORAGE}
The AFC scheme is based on spectral shaping of an inhomogeneously broadened optical transition \cite{gisin2009pra}. By applying spectral hole burning sequences, a comb with periodicity of $\Delta$ can be shaped. When a photon is injected, it will be absorbed by the comb. Its state is mapped into a collective excitation of the atoms that resonate with the photon, and can be written as a collective Dicke state: $\Sigma_j e^{-i2\pi\delta_{j} t} \left | g_1 \dots e_j \dots g_N \right \rangle$, $\delta_j= m_j\Delta $ ($m_j$ is an integer number) is the frequency detuning of the atom j \cite{reid2018optica}\cite{gisin2009pra}. The atoms will dephase since different atom will acquire a different phase ($e^{i\delta_j}t$) after a period of $t$. However, after a predetermined time $\frac{1}{\Delta}$, they will collectively rephase, leading to a photon re-emission. In order to achieve long storage time and on-demand readout, a pair of control pulses can be applied to transfer the collective excitation in and out another ground spin state. This is the so-called spin-wave AFC storage.

\begin{figure}[htbp]
\centering
\fbox{\includegraphics[width=0.95\linewidth]
{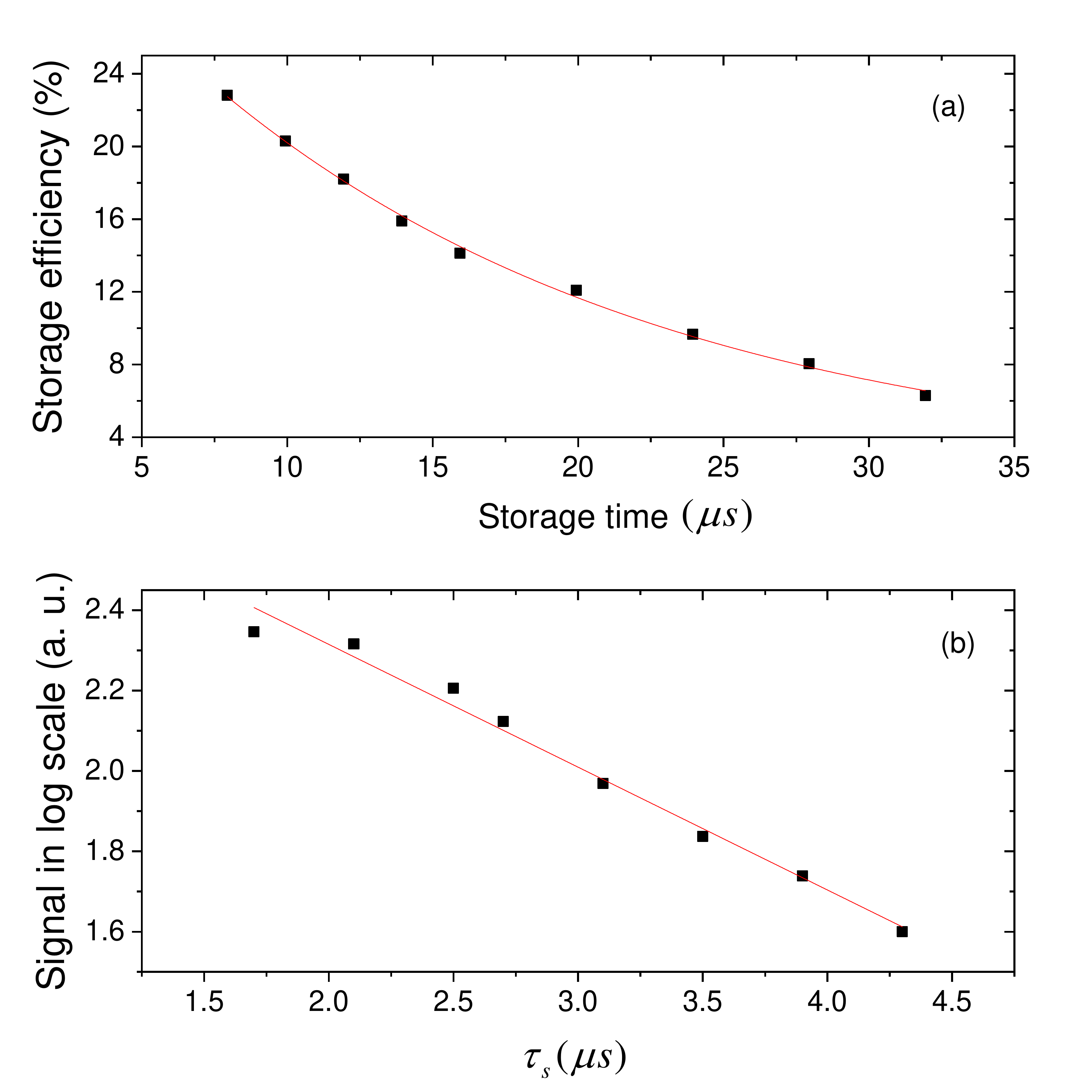}}
\caption{(a) Storage efficiency as a function of storage time for the ROSE scheme. The maximum storage efficiency is deduced as $\eta_0=34.4\ \%$ at $\tau=0$, and the extracted effective coherence time is $T_{2eff}=37.4\pm{0.9}$ $\mu$s. (b) Spin-wave AFC echo amplitude as a function of $\tau_s$. $\tau_s$ denotes the pulse spacing between the two control pulses, which is the spin-wave storage time. $T_2^*$ is fitted as $3.3\pm{0.2}$ $\mu$s. }
\label{fig.T2eff}
\end{figure}

After finishing class cleaning and spin polarization, comb on ${\left | \pm {1/2} \right \rangle}_g$ $\rightarrow$ ${\left | \pm {5/2} \right \rangle}_e$ transition can be created. Here we employ the parallel preparation sequence \cite{afz2016pra_higlymulti} to create all absorption teeth in parallel. We prepare an AFC with bandwidth of 2 MHz and a two-level storage time of 8 us. The preparation pulse has a duration of 2 ms and is repeated for 50 times to create a high quality comb. The time sequence for AFC storage is shown in Fig. \ref{fig.energy_level}. (c). The input pulse is same as previous and is injected at $t=0$. The AFC echo comes out at $t=8\ \mu s$, and the corresponding storage efficiency is $5\ \%$. In order to realize spin-wave AFC storage, two control pulses with the center frequency of $f_+$ and a chirp bandwidth of 2 MHz after the input pulse are injected. They have a duration of 1.7 $\mu$s and a peak power of 11 mW before the cryostat. The pulse spacing between them is $\tau_s$. The spin-wave AFC echo comes out at $\tau_s+8\ \mu$s and its intensity is decreased by approximately an order compared with the AFC echo. The fast dephasing on the spin state is the primary source for the reduction of efficiency. As shown in Fig. \ref{fig.T2eff} (b), $T_2^*=3.3 \pm 0.2$ $\mu$s. {The inhomogeneous spin line-width ($\gamma_{inh}$) is deduced as 114 kHz, using the formula $\gamma_{inh}=\frac{\sqrt{2ln{2}}/\pi}{T_2^*}$ \cite{reid2013njp}\cite{gisin2012njp}. As a comparison, the inhomogeneous spin line-width is 60 kHz in the bulk, measured by Raman-Heterodyne technique. This broadening magnitude is compatible with that of the optical inhomogeneous broadening.}

In order to verify the fidelity of the spin-wave AFC memory, a reference pulse with a relative phase of $\Delta \varphi '$ to input pulse is injected at $t=9.7$ $\mu$s ($\tau_s$ is fixed as 1.7 $\mu$s) to interfere with the spin-wave AFC echo. The relative phase increases with a step of 20 degrees. The intensities of the output signals are recorded by a PMT. Usually, the PMT has a very nonlinear response to the intensity of the input pulse (see supplementary material for detail), data shown in Fig. \ref{fig.interference} (b) are calibrated and normalized. {The processes on interference patterns are similar with the ROSE scheme. The interference visibility of $V=0.99 \pm{0.03}$ is obtained, revealing the phase-preserving property of the the spin-wave AFC memory.}

\section{ DISCUSSION AND CONCLUSION}
In conclusion, SMF-compatible waveguides in a $\mathrm {^{151}Eu^{3+}:Y_2SiO_5}$ crystal are fabricated using FLM. On-demand light storage is demonstrated in a waveguide based on the spin-wave AFC and ROSE scheme. The reliability of waveguide-based optical memory is demonstrated by performing a series of interference experiments based on these two schemes with almost unit visibility. {In our current waveguide configuration, the free induction decay is the main source of noise for both memory schemes. This kind of noise can be rejected by utilizing polarization filtering based on other types of waveguides which support more than one polarization mode \cite{cf2014lpr}.}

Also, we notice that the fabrication process causes absorption reduction and fast dephasing on spin states, which are unfavorable for optical quantum memory schemes, especially for the spin-wave AFC scheme. {The first problem is attributed to the expansion of the optical inhomogeneous broadening after fabrication, and can be solved by using a longer crystal or increasing the dopant concentration. The second problem is obviously correlated with the optical inhomogeneous broadening, in view of similar magnitude of broadening after fabrication. The fabrication parameters should be further optimized to find a trade-off between the single-mode coupling efficiency (preferring severer damage) and the line-width broadening (preferring weaker damage).} We note that the storage time can be further extended with the help of a carefully aligned magnetic field \cite{nature_6hour}\cite{ZEFOZ2005prl} and the dynamical decoupling sequences \cite{afz2015prl} \cite{gold2013prl}, to fulfill the requirements for long-distance entanglement distribution \cite{gisin2011rmp}.

\section*{Funding Information}
This work was supported by the National Key R\&D Program of China (No. 2017YFA0304100), the National Natural Science Foundation of China (Nos. 11774331, 11774335, 11504362, 11821404, 11654002), Anhui Initiative in Quantum Information Technologies (No. AHY020100), Key Research Program of Frontier Sciences, CAS (No. QYZDY-SSW-SLH003), Science Foundation of the CAS (No. ZDRW-XH-2019-1), the Fundamental Research Funds for the Central Universities (No. WK2470000026).

\section*{Acknowledgments}

The authors thanks the technical help from Titas Gertus and Orestas Ulcinas.

\section*{Disclosures}
\noindent\textbf{Disclosures.} The authors declare no conflicts of interest.

\section*{Supplemental Documents}
Please see Supplementary material for supporting content.






\bibliography{WG}

\bibliographyfullrefs{WG}


\end{document}